\documentstyle{mn}
\def\muas{\hbox{$\,\mu\rm as$}}
\def\mas{\hbox{$\,\rm mas$}}
\def\arcspt{{\buildrel{\prime\prime}\over .}}
\newif\ifAMStwofonts
\AMStwofontstrue
\makeatletter
\def\gsim{\compoundrel>\over\sim}
\def\lsim{\compoundrel<\over\sim}
\def\compoundrel#1\over#2{\mathpalette\compoundreL{{#1}\over{#2}}}
\def\compoundreL#1#2{\compoundREL#1#2}
\def\compoundREL#1#2\over#3{\mathrel
      {\vcenter{\hbox{$\m@th\buildrel{#1#2}\over{#1#3}$}}}}
\makeatother
\ifoldfss
  \ifCUPmtlplainloaded \else
    \NewTextAlphabet{textbfit} {cmbxti10} {}
    \NewTextAlphabet{textbfss} {cmssbx10} {}
    \NewMathAlphabet{mathbfit} {cmbxti10} {} 
    \NewMathAlphabet{mathbfss} {cmssbx10} {} 
  \fi
  \ifAMStwofonts
    \ifCUPmtlplainloaded \else
      \NewSymbolFont{upmath} {eurm10}
      \NewSymbolFont{AMSa} {msam10}
      \NewMathSymbol{\upi}     {0}{upmath}{19}
      \NewMathSymbol{\umu}     {0}{upmath}{16}
      \NewMathSymbol{\upartial}{0}{upmath}{40}
      \NewMathSymbol{\leqslant}{3}{AMSa}{36}
      \NewMathSymbol{\geqslant}{3}{AMSa}{3E}

       \let\le=\leqslant
       \let\ge=\geqslant
    \fi
  \fi
\fi 
\ifnfssone
  \newmathalphabet{\mathit}
  \addtoversion{normal}{\mathit}{cmr}{m}{it}
  \addtoversion{bold}{\mathit}{cmr}{bx}{it}
  \newmathalphabet{\mathbfit} 
  \addtoversion{normal}{\mathbfit}{cmr}{bx}{it}
  \addtoversion{bold}{\mathbfit}{cmr}{bx}{it}
  \newmathalphabet{\mathbfss} 
  \addtoversion{normal}{\mathbfss}{cmss}{bx}{n}
  \addtoversion{bold}{\mathbfss}{cmss}{bx}{n}
  \ifAMStwofonts
    \ifCUPmtlplainloaded \else
      \UseAMStwoboldmath
      \makeatletter
      \new@mathgroup\upmath@group
      \define@mathgroup\mv@normal\upmath@group{eur}{m}{n}
      \define@mathgroup\mv@bold\upmath@group{eur}{b}{n}
      \edef\UPM{\hexnumber\upmath@group}
      \new@mathgroup\amsa@group
      \define@mathgroup\mv@normal\amsa@group{msa}{m}{n}
      \define@mathgroup\mv@bold\amsa@group{msa}{m}{n}
      \edef\AMSa{\hexnumber\amsa@group}
      \makeatother
      \mathchardef\upi="0\UPM19
      \mathchardef\umu="0\UPM16
      \mathchardef\upartial="0\UPM40
      \mathchardef\leqslant="3\AMSa36
      \mathchardef\geqslant="3\AMSa3E

       \let\le=\leqslant
       \let\ge=\geqslant
    \fi
  \fi
\fi 
\ifnfsstwo
  \DeclareMathAlphabet{\mathbfit}{OT1}{cmr}{bx}{it}
  \SetMathAlphabet\mathbfit{bold}{OT1}{cmr}{bx}{it}
  \DeclareMathAlphabet{\mathbfss}{OT1}{cmss}{bx}{n}
  \SetMathAlphabet\mathbfss{bold}{OT1}{cmss}{bx}{n}
  \ifAMStwofonts
    \ifCUPmtlplainloaded \else
      \DeclareSymbolFont{UPM}{U}{eur}{m}{n}
      \SetSymbolFont{UPM}{bold}{U}{eur}{b}{n}
      \DeclareSymbolFont{AMSa}{U}{msa}{m}{n}
      \DeclareMathSymbol{\upi}{0}{UPM}{"19}
      \DeclareMathSymbol{\umu}{0}{UPM}{"16}
      \DeclareMathSymbol{\upartial}{0}{UPM}{"40}
      \DeclareMathSymbol{\leqslant}{3}{AMSa}{"36}
      \DeclareMathSymbol{\geqslant}{3}{AMSa}{"3E}

       \let\le=\leqslant
       \let\ge=\geqslant
    \fi
  \fi
\fi 
\ifCUPmtlplainloaded \else
  \ifAMStwofonts \else 
    \def\upi{\pi}
    \def\umu{\mu}
    \def\upartial{\partial}
  \fi
\fi

\title{Space-borne global astrometric surveys: the hunt for extra-solar
planets.}
\author[M.G. Lattanzi et al.]{M.G. Lattanzi,$^1$
     A. Spagna,$^1$  A. Sozzetti,$^{1,2}$
    S. Casertano,$^2$\thanks{Affiliated to the Space Sciences Dept., ESA}\\
    $^1$Osservatorio Astronomico di Torino, Pino Torinese TO, I-10025, Italy\\
    $^2$Space Telescope Science Institute, Baltimore MD, 21218, USA}
\date{Accepted . Received}

\pagerange{\pageref{firstpage}--\pageref{lastpage}}

\pubyear{2000}
\begin{document}
\maketitle
\label{firstpage}

\begin{abstract}

The proposed global astrometry mission {\it GAIA}, recently recommended
within the context of ESA's Horizon 2000 Plus long-term scientific
program, appears capable of surveying the solar neighborhood within
$\sim$ 200 pc for the astrometric signatures of planets around
stars down to the magnitude limit of $V$=17 mag, which includes 
late M dwarfs at 100 pc.

Realistic end-to-end simulations of the GAIA global astrometric measurements
have yielded first quantitative estimates of the sensitivity to
planetary perturbations and of the ability to measure their orbital
parameters. Single Jupiter-mass planets around normal solar-type 
stars appear detectable up to 150 pc ($V\le $12 mag) with
probabilities $\ge$ 50 per cent for orbital periods between $\sim$2.5 and
$\sim$8 years, and their orbital parameters measured with better than
30 per cent accuracy to about 100 pc.  Jupiter-like objects (same mass and
period as our giant planet) are found with similar probabilities up to
100 pc.

These first experiments indicate that the {\it GAIA} results
would constitute an important addition to those which will come from 
the other ongoing and planned planet-search programs. These data 
combined would provide a formidable testing ground
on which to confront theories of planetary
formation and evolution.

\end{abstract}

\begin{keywords}

astrometry -- stars:planetary systems --
artificial satellites,space probes
\end{keywords}

\section{Introduction}

At the very end of 1995, the discovery~\cite{mayo} of the first
Jupiter-mass ($M_J$) planet orbiting a normal star other than the Sun heralded
the beginning of a new era of extraordinary discoveries in the realm of
extra-solar planets, bringing with them the hope for a better
understanding of the formation and frequency of planetary systems, and
perhaps of bringing us closer to the ultimate goal of discovering
extraterrestrial life.

After four years since that discovery, spectroscopic programs have been
able to reveal some twenty extra-solar planets, i.e., objects with a lower
mass limit below the 13-$M_J$ cut-off which has been adopted by
Oppenheimer and Kulkarni~\shortcite{oppen99} 
to differentiate giant planets from brown dwarfs.

However, these discoveries have raised new and troubling questions
in our understanding of the properties of planetary systems.
The fundamental tenets upon which present theories are
based include nearly circular orbits and giant planets formed
several AU from the central star, in contrast with the very short
orbital periods~\cite{mayo} and high
eccentricities~\cite{lath89,coch,mazeh}  found for several of
the new discoveries. Their interpretation as {\it bona-fide} 
planets rests on our understanding of correlations shown by their orbital
and physical parameters, as recently discussed by Black~\shortcite{black} 
and earlier by Duquennoy \& Mayor~\shortcite{duque} 
in their work on solar-type binary stars.

New models, which employ specific physical and dynamical
mechanisms like {\it in-situ} formation~\cite{bode2} or orbital
migration~\cite{lin1,trill,murray}, have been proposed to justify the presence 
of {\it hot jupiters} around normal stars, demonstrating that
the interplay between additional theoretical
work and more observational data will be necessary for a continued
improvement in our theoretical understanding of how planets form and
evolve, and where Earth-like planets could eventually be found.

However, simply adding a few tens of additional detections of giant
extra-solar planets is not enough.  A better understanding of the
conditions under which planetary systems form and of their general
properties requires large, {\it complete} samples of planets, with
useful upper limits on Jupiter-mass planets at several AU from the
central star.\\   
Ongoing and planned radial velocity surveys~\cite{mayo,coch,noyes,marcy,marcy2}
have started filling significant portions of the relevant parameter space.
Searches based on relative astrometry from the ground and in 
space~\cite{map,pti,keck,vlti,pravdo} 
will be an important complement to the spectroscopic
work and, probably, the preferred means for establishing the existence of 
planets around young stars and that of low mass planets down to a few
Earth masses, as will be the case for SIM~\cite{boden,unwin}.

A {\it HIPPARCOS}-like, space-borne global
astrometric mission, which can survey the whole sky to
faint magnitudes and with high astrometric accuracy, will enable the 
monitoring of large ($> 10^5$) samples of stars, with well
understood completeness properties. This, depending on actual values of 
planetary frequencies~\cite{ppiv}, could yield the possibility of
making firm measurements of
statistical properties of planetary systems. For,
different correlations among orbital parameters (eccentricity, period or 
semi-major axis) and measurable differences in planetary frequency are likely 
to be generated by diverse planetary formation scenarios (core accretion and 
disk instability are the two known to date) and evolution mechanisms, 
as well as different formation and evolution processes of the parent star 
(binarity, spectral type, metallicity, age). 
An astrometric mission such as {\it GAIA} appears well poised
for such a systematic census of planetary systems within $\sim$ 200 pc
from the Sun.

The {\it GAIA} concept was originally proposed by Lindegren and
Perryman~\shortcite{linde} as a possible 
Cornerstone--class mission within the Horizon 2000+ program of
scientific satellites of the European Space Agency. This satellite
is designed to chart more than one billion objects (stars,
extra-galactic objects, and solar system objects) on the sky down
to the limiting magnitude of $I=20$. The targeted final accuracy is $\sim$
10 $\mu$as on positions and parallaxes, and $\sim$ 10
\muas/year on proper motions at the reference magnitude of
$V=15$ for a G2V star~\cite{gilm}, and for a mission life time of 5 years.

In the following sections we show and discuss some relevant results
derived from detailed end-to-end simulations of the data acquisition and
analysis process for {\it GAIA},
which, as we will see, appears capable of discovering Jupiter-mass
planets around $\sim$ 3$\times 10^5$ candidate stars (including dwarfs 
earlier than K5).

\section{Data simulation}

The simulation code is an adaptation of that used by Galligani et
al.~\shortcite{galli} for the assessment of the astrometric accuracy of
the sphere reconstruction in the {\it HIPPARCOS} mission.  We generate
catalogs of single stars randomly distributed on the sky; each run
produces a sphere of 200 stars. As the satellite observing strategy (or 
scanning law) is most easily described in ecliptic coordinates, positions, 
proper motions and parallax ($\lambda$, $\beta$, $\mu_\lambda$, $\mu_\beta$, 
and $\pi$, respectively) are also given in the same coordinate system. 
For the moment, $\mu_\lambda$, $\mu_\beta$ and $\pi$, 
as well as magnitudes and colors, are drawn from simple
distributions which do not represent any specific Galaxy model; in
particular, in each run the 200 stars simulated have the {\it same}
parallax, total proper motions, magnitudes, and colors.

The satellite is made to sweep the sky in such a way that
the spin axis precesses around the Sun at a rate of about
6.5 rev/year and with a constant angle of $ 43 \deg $.  Stars that at
any given time are ``seen'' within a strip $ \sim 1\deg $ wide along the great
circle (GC) being scanned are considered observed; a GC is
completed in about 2.5 hours.  Basic observations, which in principle
can easily be derived from the fringe pattern measurements above, are
the abscissae along a GC, as {\it GAIA}, like {\it HIPPARCOS},
makes very precise measurements in one dimension only.  The mission
lifetime is set to 5 years and the scanning law is such that the number
of basic observations per star is a function of ecliptic latitude: each 
given object on the sky is re-observed approximately every month, for a total
of $\sim 60$ one-dimensional position measurements, on average. The minimum 
number of observations is $\sim$ 30 and occurs for stars at the ecliptic 
equator. The position of a star at a given time, as described by the 
combined effects of parallax and proper motions, is called here {\it barycentric}
location and it has been described in Euclidean space (flat Lorentzian).
General relativistic effects, which will have to be considered in the
future (especially for the case of Earth-like planets), are not taken
into account.

Finally, gravitational perturbations (Keplerian motions) induced by a
({\it single}) nearby orbiting mass are added to the barycentric
location resulting in the ``true'' {\it geometric} location of a target.

Simulated observations are generated by adding the appropriate astrometric
noise to the true locations.  The error sources considered in the
simulation are discussed below.

As in {\it HIPPARCOS}, {\it GAIA} will have two viewing directions separated by a
large angle named Base Angle (BA).  It is by connecting directions far
apart on the sky that the principle of space-borne global astrometry
demonstrated by {\it HIPPARCOS} is implemented.  Therefore, it is not
surprising that the accuracy with which the BA is known throughout the
mission is probably the most important single item within the {\it GAIA}
concept.  At the moment there are two mature optical designs for the BA.
One configuration feature two telescopes mounted on the same optical
bench and pointing along the two different line-of-sights.  The other
design uses a beam combiner (an adaptation of that used on {\it HIPPARCOS})
which physically materializes the BA and multiplexes the two viewing
directions into a single telescope unit; beam combiner and telescope are
bolted to the same bench.  The twin
telescopes designed for the first configuration feature an off-axis
monolithic primary (with no central obscuration), while the collecting
telescope of the latter design is a Fizeau
interferometer with a baseline of 2.45m.

The details of the {\it GAIA} optical configuration (see e.g.
Gilmore et al.~\shortcite{gilm}, and references therein) 
are not critical to our discussion
of photon-driven astrometric errors.  The major difference with our
earlier work~\cite{case} is that both designs feature a significantly
increased telescope aperture.  The monolithic configuration has a
rectangular primary of 1.7m by 0.7m, and each circular aperture of the
interferometric option is 0.65m in diameter.  High-accuracy measurements
of the position (phase) of the PSFs are made directly on the focal plane
using CCD detectors (see next section).

\begin{table}
\caption{{\em Photon and total error for a single observation. This is
the combination of $\simeq$ 8 consecutive field-of-view crossings, for any
given direction on the sky, each comprising 20
elementary 1-sec exposures, for a total duration for the single
integration of $\simeq$ 160 sec. These values strictly apply to the
interferometric option discussed in the text and to a G2V star and
negligible interstellar absorption (Av = 0). The pure photon noise values
increase by $\sim$
30 \% for an A0V star and improve by $\sim$ 20 \% for a K3V. 
 More explanations are provided in
Sections 2.2-2.3}}
\label{errors}
\catcode`\!\active \def!{\hbox{\phantom{0}}}
\begin{tabular}{@{}cccc}
$ V $ mag & Photon error
&\multicolumn{2}{c}{Total error $\sigma_\psi$ ({\muas})}\\
 & ({\muas}) & ($\sigma_{\rm b}$ = 30 pm) & ($\sigma_{\rm b}$ = 10 pm) \\
 10   &   !!2.4!!   &   !!10.2   &   !!3.8! \\
 12   &   !!5.9!!   &   !!11.5  &   !!6.6! \\
 15   &   !24.6!!   &   !26.5   &   !24.8! \\
 17   &   !70.3!!   &   !71.0   &   !70.4! \\
 18   &   129.9!!   &   130.3   &   129.9! \\
\end{tabular}
\end{table}

\subsection{Payload configurations}

The payload design has greatly advanced since the idea sketched in
Lindegren \& Perryman~\shortcite{linde} and further developed in
Loiseau \& Shaklan~\shortcite{loise}.  
The two industrial studies commissioned by ESA 
have looked into the feasibility of two
different options for the {\it GAIA} payload~\cite{gilm}: an all-passive
configuration with two identical telescopes (with rectangular-shape
monolithic primary mirrors) to be operated in L2, and an all-active
configuration with an interferometric (diluted) beam combiner and a
Fizeau interferometer as light collector behind it to be operated in
geosynchronous orbit.  Stability of the BA is passively maintained in
the case of the monolithic configuration by utilizing a sophisticated
active thermal control system (which must operate at the $\mu$K level)
and a silicon carbide structure for the optical bench.  On the other
hand, the interferometric configuration is all-active, with the
stability achieved by real-time monitoring implemented via high-accuracy
laser metrology~\cite{gai97} and control of all critical degrees of
freedom.

As we are more familiar with the interferometric option for 
{\it GAIA}~\cite{latt97}, we will be referring to that in the discussion below.
However, precision and accuracy requirements and estimates are quite similar 
for both configurations, especially at the bright magnitudes we are 
concerned with in this paper.

\subsection{Photon noise and point spread function measurement}

The ability to measure accurately the position of a star ultimately
depends on the width and shape of the point spread function (PSF) of
the imaging system and on the number $ N $ of photons detected.  For a
well-sampled PSF, the theoretical limit is shown by~\cite{oldlinde} to
be $ \epsilon = \lambda / (4 \pi x_{\rm rms} \sqrt{N} ) $, where $
x_{\rm rms} $ is the rms size of the aperture in the measurement
direction.  For two circular apertures of diameter $ D $ and with a
central separation $ B $, we have $ x_{\rm rms} = \sqrt {(B/2)^2 +
(D/4)^2} $; for the baseline {\it GAIA} parameters ($ B = 2.45 $~m, $ D
= 0.65 $~m), $ x_{\rm rms} = 1.24 $~m~\cite{linper}.  For $ \lambda =
750 $~nm (the baselined operational wavelength), this translates into a
theoretical monochromatic measurement accuracy of $ 10 {\mas}/\sqrt {N}
$.

The measurement accuracy is degraded for non-monochromatic measurements,
by about a factor 2 for a Gaussian filter centered at 7500~{\AA} and
2000~{\AA} wide~\cite{gai98}.  In addition, the requirement of optimal
sampling may be difficult to achieve, since the central fringe is only
about $ 0\arcspt04 $ wide and the field of view is $ 1\deg $.  In
practice, this will probably cause a loss of accuracy of about 20--40
per cent~\cite{gai,gai98}.  In the following, we assume a ``best
reasonable'' single-measurement photon noise error of $ 24
{\mas}/\sqrt{N}$.

Since scans overlap partially, each ``observation'' of a star will
consist of about 8 consecutive scans with 20 sec of integration time
allocated per scan, a total integration time of 160~s.  Assuming a total
collecting area of 0.664~$\rm m^2 $ (2 apertures of 0.65~m diameter) and
a total system efficiency of 20 per cent, a star with $ V = 15 $ would
generate about $ 10^6 $ photons per observation, corresponding to a
photon-limited measurement accuracy of the fringe position of
24~{\muas}.  The accuracy scales with the inverse square root of the
flux.  This approximate calculation agrees with the values in 
Table~\ref{errors},
obtained for the current version of the interferometric option.  The
photon noise values (second column in Table~\ref{errors}) for  
$10\le V\le 18$ stars were obtained with 3.1 pixels per fringe period, 
a RON of 3 electrons/pixels, and a DQE of 0.6. 
Objects brighter than $V$ = 10 mag will be also observed by {\it GAIA}.
However, limitations
on the dynamic range achievable with CCDs, requires to limit the exposure
time for the brighter stars. Thus is practice, the precision starts to
level off brighter than $V\simeq$ 10.

This accuracy is based on the photon statistics only, and does not take
into account possible systematic effects, such as distortions in the
optical system or in the detector, imperfection in the fringes, and so
on.  Many such systematic effects can be calibrated using closure
methods, others will require enhancements in system design.

\subsection{Accuracy of the Base Angle}

The systematic effects mentioned above will lead to residual
(systematic) errors in the knowledge of the BA which do not scale with
magnitude.  Ultimately, these will be the errors which will limit the
maximum accuracy of {\it GAIA} for bright sources.  We combine such residual
errors in what is hereafter called {\it residual bias}, or simply {\it
bias}.  In the present error budget we assume that the bias can be
described as a stationary stochastic process; therefore the bias
contributed at the single-observation level to each object (Table~\ref{errors})
scales, like photon noise precision, with the average number of
observations collected over the mission lifetime.

The BA can be measured and monitored accurately over time intervals
longer than a full revolution ($\ge$ 3 hours) by making use of the $
2\pi $ closure properties of the consecutive scans.  This is an
especially important feature of {\it GAIA} whose potential, at the $mas$
level, has been beautifully proven by {\it HIPPARCOS}.  However, the bias over
shorter time scales need to be controlled by ensuring that the relative
positions and shapes of all optical elements of the beam combiner do not
vary throughout the observation.
 The necessary sub-nm accuracy will probably be achieved by a
combination of passive control and of laser metrology.  For simplicity,
we summarize the overall accuracy (bias) with which the (relative)
positions of the mirrors of the beam combiner must be known by a single
number, the ``baseline error'' $ \sigma_{\rm b} $.  This helps
visualize the complex interplay of the beam combiner mirrors
(which materialize the two baselines of the interferometric design) with a
more familiar bias of the type $B\times \delta$, where B is the baseline
and $\delta$ the angular uncertainty\footnote{While the main
contribution to the baseline error will probably come from the relative
position of the beam combiner mirrors, the motions and/or distortions of
other optical elements (such as the primary apertures of the Fizeau
interferometer) can also generate an effective baseline error.}.
 This also helps understand the bias contribution in Table~\ref{errors}.  
For example, if the photon error is subtracted (in quadrature) from the
total error for the case with $ \sigma_{\rm b} = 30 $ pm, the resulting
value, $\sim 10$ {\muas}, represents the angular bias corresponding to
that linear ``baseline error''.  Therefore, the requirement on the
metrology is $\simeq$ 3.8 times more stringent that one would have
guessed from the simple calculation $B \times \delta$, with B=2.45~m and
$\delta$=10 {\muas}.  This example shows that the baseline bias is
driven by the dimension of the single mirrors (D=0.65~m) forming the
baseline (B/D$\simeq$3.8) within the beam combination system.

Laser metrology in laboratory settings has already achieved extremely
high performances, with relative measurements at the picometer
level~\cite{gurs,noeck2,noeck1,rease}.  Such precision has been reached
over short (few wavelengths) variations in path length, which are
appropriate to the {\it GAIA} design if a good active thermal control is
included.

However, the few pm error quoted refers to the precision and stability
of a one-dimensional laser gauge measurement of a single optical path.
Maintaining the accuracy of the interferometer baseline is much more
complex, first, because the three-dimensional positions of several
optical elements may need to be monitored simultaneously, and second,
because of the possible differences between the optical path of
starlight and of the laser gauge beams.  Noecker~\shortcite{noeck1}
lists a number of possible systematic errors for the {\it POINTS}
mission concept.  An experiment to demonstrate picometer laser metrology
for a three-dimensional system on the {\it GAIA} scale is underway~\cite{gai97}.
For the moment, it remains difficult to give firm figures for the
baseline accuracy that will eventually be achieved.  In Table~\ref{errors}, 
we consider two cases which probably bracket realistic expectations: a more
``conservative'' accuracy $ \sigma_{\rm b} = 30 $~pm, which already
appears within reach from the preliminary results available, and an
``optimistic'' accuracy $ \sigma_{\rm b} = 10 $~pm.  Notice that these
bias values are intended at the level of what is called here the
single-observation error.  The accuracy levels at the end of the mission
will be $\sim 3$ {\muas} and 1 {\muas} for $\sigma_{\rm b}=30$~pm and
10~pm, respectively, if errors in successive passes are uncorrelated, as
discussed previously. Notice that, for $\sigma_{\rm b}=30$~pm, the astrometric 
precision begins to level off brighter than $V = 12$.

Finally, it is important to mention that the numbers quoted for the linear
bias apply to a 3-dimensional monitoring of the relevant structure. The laser
beams will have to monitor the corresponding one-dimensional degrees of
freedom with significantly better resolution---a factor $<2$ for the
current design.

\section{DETECTION AND ORBIT DETERMINATION METHODS}

The magnitude of the gravitational perturbations induced by a planet on
the parent star, as seen by an astrometric mission, can be quantified
through its {\em astrometric signature} $\alpha$ defined as:
\begin{equation}
\alpha = \frac{M_p}{M_s}\frac{a_p}{d}
\end{equation}
where $M_p$, $M_s$ are the masses of the planet and star respectively, $a_p$
the semi-major axis of the planetary orbit, $d$ the distance of the system from
us. If $a_p$ is in AU and $d$ in parsec, then $\alpha$ is expressed in arcsec.

{\it GAIA}'s sensitivity to this signal depends of course on the errors
$ \sigma_\psi $ of each measurement, with theoretical
values listed in Table~\ref{errors}.  We have verified through
tests with different values of $ \sigma_\psi $ (see section~\ref{detect}) 
that, as could be expected, the detection probability depends on
$ \sigma_\psi $ and $ \alpha $ only through their ratio,
the ``astrometric'' {\em signal-to-noise ratio}
 \begin{equation}
 S/N = \alpha / \sigma_\psi,
 \end{equation}
 so that the results obtained can be straightforwardly rescaled to
different measurement errors.

For simplicity, we have thus kept the single-observation error
fixed at $ \sigma_\psi =
10{\muas} $ throughout our simulations.  This is the value expected
for
relatively bright stars ($V\le$ 12 mag, corresponding to the Sun at 200
pc), with a conservative baseline error of 30 pm (see
Table~\ref{errors}).  Our simulations should thus give realistic
estimates of: {\em a)} {\it GAIA}'s detectability horizon of planetary
mass companions to solar-type stars in the vicinity of our Sun,
and {\em b)} limits on distance for accurate orbital parameters
determination.

The starting point for our two-level statistical investigation is the
computation of detection probabilities, which will in principle depend
upon 1) mission parameters, 2) noise sources, and 3) orbital elements.
The contributions to points 1) and 2) are listed in Section 2.  As for
point 3), we will express the detection probabilities as function of
the period $P$ and the signature $\alpha$, which we expect to be the
major contributors, and generally average over the expected distribution
of the other orbital parameters.

\subsection{Detection}

Our first detection method applies a standard $\chi^2$ test (with the
confidence level set to 95 per cent) to the residuals $\psi - \psi_r$,
where the $\psi$ are the actual measurements, and the $\psi_r$ are the
GC abscissae recomputed after fitting the observations of each
star with a single-star model, i.e., solving only for the five
astrometric parameters appropriate for a single star (position,
parallax, and proper motion).
Being $\chi^2_o$ the value provided by the null model (no planet),
the test fails when Pr$(\chi^2 \le \chi^2_o) \ge 0.95$.
In this case the planet is considered {\em detected}.
Note that this method only measures deviations from the single-star model,
it makes no
assumptions on the nature of the residuals nor does it provide initial
guesses for the computation of the planet's mass and orbital parameters.

We employed the $\chi^2$ test to analyze 160\,000 stars uniformly
distributed on the sky, perturbed by dark companions inducing
astrometric signatures $\alpha$ ranging from 10 to 100 {\muas}, with
periods P between 0.5 and 20 years.  The remaining orbital elements were
distributed randomly in the ranges: $0^\circ\le i\le 90^\circ$, $0\le
e\le 1$, $0\le \Omega\le 2\pi$, $0\le \omega\le 2\pi$, $0\le \tau\le
P$.  The single observation error was set to $\sigma_\psi$ = 10 $\mu$as,
thus the signal-to-noise ratio varied in the range $1\le S/N \le 10$.
Finally, we repeated the same simulations without planets, to verify the
correct behaviour of the test and the choice of the confidence level.
As expected the number of {\it false} detections was $\simeq$ 5 per
cent.

\subsection{Orbital parameters}

Once a planet is ``detected'', the goal is to derive reliable estimates
for of its orbital elements and mass. 
For a complete reconstruction of the orbital geometry,
we implemented an analytic model in which the observation
residuals $\psi-\psi_r$ are evaluated with a recomputed abscissa of the form:
\begin{equation}
\psi_r =
\psi_r(\lambda,\,\beta,\,\mu_\lambda,\,\mu_\beta,\,\pi,\,T_1,\,T_2,\,T_3,\,T_4,\,
e,\,P,\,\tau)
\end{equation}

The 1-d measurement along the scanned GC is thus expressed as function
of both the standard astrometric parameters
($\lambda$, $\beta$, $\mu_\lambda$, $\mu_\beta$, $\pi$) and the parameters 
describing the stellar relative orbit with respect to the barycenter of the 
planetary system:
period $P$, periastron epoch $\tau$, eccentricity $e$ and the 4 Thiele-Innes
elements $T_i$, functions of semi-major axis $a$, inclination $i$, argument of
periastron $\omega$, position angle of the ascending node $\Omega$.

The solution of the non linear system of observation equations is obtained
employing an iterative linear Least Squares procedure
by means of which the entire set of orbital elements can be simultaneously
estimated within a well defined accuracy level. The details of this procedure
are given in the following sections.

\section{RESULTS}

In this section we account for what is our present
understanding of {\it GAIA}'s sensitivity to astrometric perturbations 
induced by Jupiter-mass planets orbiting around nearby stars. 
Given the present
evidence of a number of extra-solar planets, we also present results 
on how {\it GAIA} would perform on a selection of three such systems.

\subsection{Detection probabilities}\label{detect}

\begin{figure}
\vspace{6.5cm}
\includegraphics{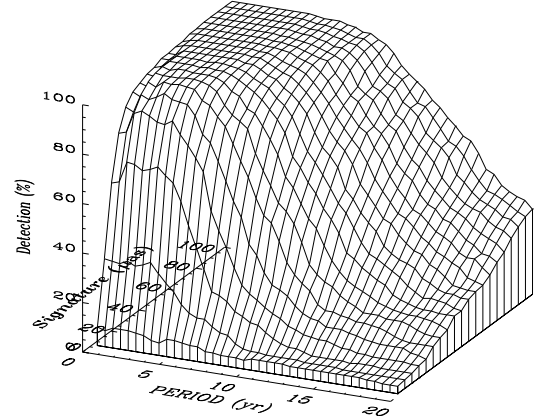}
\caption{\em
   Planet detection probability (percentage of failures of the $\chi^2$ test)
as function of the astrometric signature
   $\alpha$ and of the orbital period $P$, for $\sigma_\psi$ = 10 {\muas}.
   The percentage of detection of each point is based on 200 random planetary
   systems uniformly distributed on the sky.}
\label{tred}
\end{figure}

Figure~\ref{tred} gives the percentage of failure of the $\chi^2$ test on the
single-star hypothesis in the case of $\sigma_\psi=10$ {\muas}.
We note that at relatively low $S/N$ ratios the detection probability is
dominated by sampling of the orbital period. At higher $S/N$ values
orbital sampling is less critical and long period planets (up to about twice
the mission duration) are detectable: as a matter of fact, when $S/N
\rightarrow$ 10 detection probability reaches about 100 per cent.

\begin{figure}
\vspace{6cm}
\includegraphics{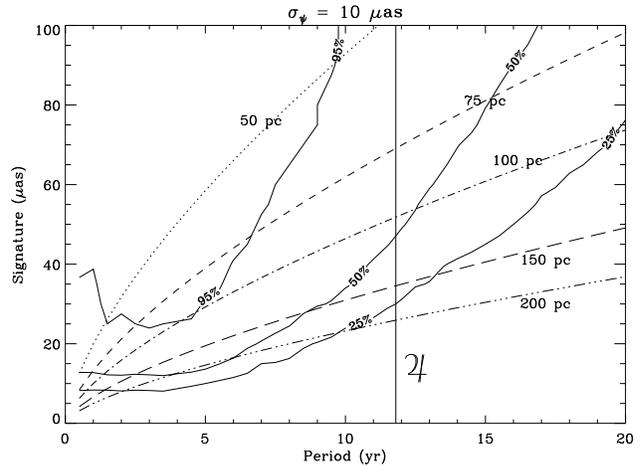}
\caption{\em Iso-probability contours ({\it solid lines}) for 25
per cent, 50 per cent and 95 per cent of detection probability,
compared with Kepler's third laws ({\it dotted/dashed lines}) for
systems with Jupiter-Sun masses at D=50, 75, 100, 150 and 200 pc.
Jupiter-like planets ($P=11.8$ yr) appear detectable, with
probability $\ge$ 50 per cent, up to a distance of 100 pc
(vertical line).} \label{sp50}
\end{figure}

The dip at $P = 1$ year was somewhat expected, as both orbital motion and
parallactic factor have the same period.  However, the decrease is
small, indicating that the coupling is less critical than might have
been anticipated.  This is probably due to the fact that the parallactic
motion has fixed phase and aspect ratio, and a relatively small mismatch
in any of the orbital parameters---phase, inclination, eccentricity---is
sufficient to separate the two signals.

Figure~\ref{tred} also shows that the $ \chi^2 $ test quickly loses its
sensitivity for $S/N$ approaching unity; thus planets for which the
error in individual observations is comparable to the apparent semi-major
axis are essentially undetectable with this technique.

As mentioned before, the results of Figure~\ref{tred} apply to
measurement errors other than the assumed $ \sigma_\psi = 10$ {\muas},
as long as the $S/N $ value remains the same.  For example, for a single
observation error $\sigma_\psi = 1$ {\muas}, the detection probability
is exactly the same as shown in Figure~\ref{tred}, but for amplitudes
ten time smaller.

We can get important physical informations from the statistical results
of Figure~\ref{tred}, simply changing the interpretative perspective
from which we are leading the discussion.  The solid lines in
Figure~\ref{sp50} are the empirical relations, derived from
Figure~\ref{tred}, that express the amplitude of the perturbation as a
function of orbital period needed for detection probabilities of 25, 50,
and 95 per cent, respectively.  If the orbital period is shorter than
the assumed mission life-time of 5 years, the detection probability is
about 50 per cent for $ S/N \sim 1 $.  For longer periods, the sampling
of the orbit is worse, probability drops significantly and a much higher
signal is required for the planet's signature to be detected.  This is
in qualitative agreement with the results of Babcock~\shortcite{bab}, who
studied the {\em detection and convergence probability} of a complete
orbital model for simulated planetary systems as observed by the mission
{\it {\it POINTS}}.  Babcock did find a slightly larger sensitivity on
planet's period, manifested in an earlier turn-up and steeper slope at
long periods of the 50 per cent probability curve; this most likely
depends on the fact that the determination of reliable orbital elements
is more challenging than detection only.

Let us now discuss the other elements of Figure~\ref{sp50}.  The
overplotted dashed and dotted lines represent the signal expected for a
Jupiter-mass planet and solar-mass star at various distances and orbital
periods, obtained by substituting Kepler's third law in the expression
for $ \alpha $:

\begin{equation}
\alpha\simeq \frac{M_p}{M_s^{2/3}}\frac{P^{2/3}}{d}
\label{kepler}
\end{equation}

The vertical solid line at $P$ = 11.8 years indicates the locus
of the actual Jupiter-Sun system at different distances.  Jupiter-Sun
systems appear detectable with probability $\ge 50$ per cent up to a
distance of 100 pc, while Jupiter-mass planets with shorter periods can
be detected to larger distances: to 150 pc for periods between 2.5
and 8 years. The relation can be rescaled to lower planet masses by
reducing the distance of the system in proportion to the planet mass, since
it is the ratio  $M_p / d$ that enters in the astrometric signature $\alpha$.

Periodic photospheric activity  across the disk of the parent star 
(e.g. star-spots cycles)
can induce displacements in the position of the photocenter, thus adding 
astrometric signal which could be difficult to disentangle from the planet 
signature. However, the magnitude of such intrinsic astrometric noise appears
to be at most of a few $\mu$as~\cite{woolf} for a solar-type 
star at a distance of 10 pc, i.e., significantly smaller than the expected 
single-measurement errors.

Finally, although the single-observation error significantly deteriorates
with magnitude (Table~\ref{errors}), 
jovian planets around late type stars can reliably be detected at
relatively large distances.

\subsection{Orbital parameters estimation}

The next step beyond the simple detection of candidate planets is the
task of estimating the orbital parameters of each system and
the mass of the unseen companion.
\begin{figure}
\vspace{11cm}
\includegraphics{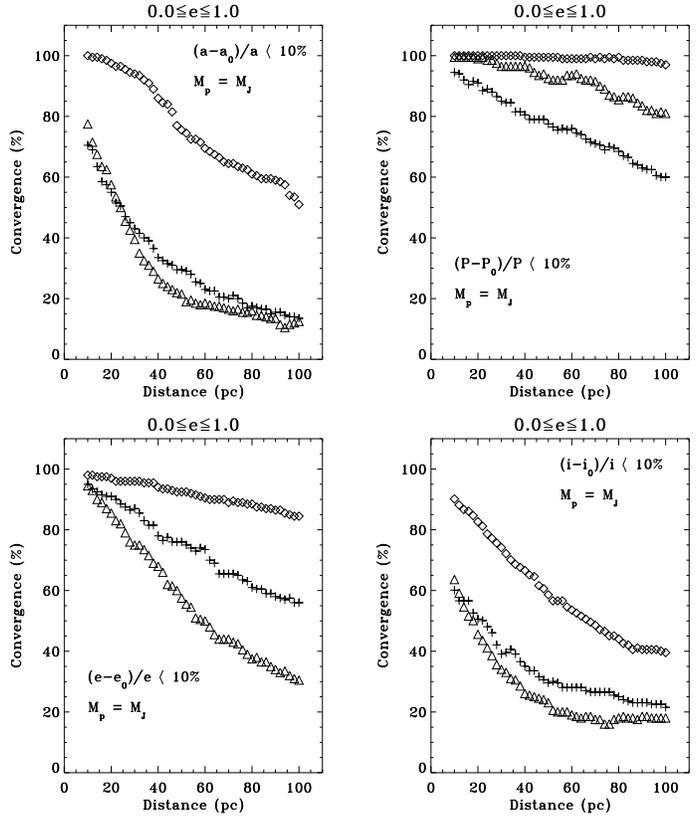}
\caption{{\em Convergence probability to a 10 per cent accuracy for $a$, $P$, $e$, $i$,
as function of the distance from the Sun.}}
\label{conv}
\end{figure}

Our method, described in more detail below, consists of applying an
iterative non-linear least-squares fitting procedure to each of the simulated 
orbits; the 'known' orbital parameters are utilized as initial guesses to start 
the fit. 
Convergence of the non-linear fitting method and quality of
the orbital solutions can both be significantly affected by the choice
of the starting guesses. 
Therefore, the  use of the true values of the orbital parameters to 
initialize the fit leaves open some important
questions related to how and to what extent effective starting values, i.e., 
leading to  successful orbital solutions,  will be identified from the 
data as function of actual performances of the satellite, uncertainties in the error
model, and properties intrinsic to planetary systems. For this, realistic global 
search strategies must be implemented and double-blind test campaigns
conducted. Work on these issues is in progress and will be reported in the
future. \\
 Instead, 
this work focuses on the important goal of 
gauging GAIA's {\it ultimate} ability in measuring planets 
(properly, single giant planets orbiting single solar-type stars). 
An efficient way of achieving this is by assuming perfect knowledge 
of GAIA's characteristics (mainly, error model and satellite attitude) 
and, indeed, 
by using the known values of each orbital parameter as suitable initial guesses
for the least-squares solutions 
\footnote{Note that, as explained later in this section, 
the fitting program does not know that the initial values 
provided for the parameters are also their true values}. 
Then, the post-fit differences to the true values of the
parameters should be a reliable measure of the accuracy in orbit reconstruction
that can ultimately be achieved by GAIA with the given measurement errors.

\begin{figure}
\vspace{11cm}
\includegraphics{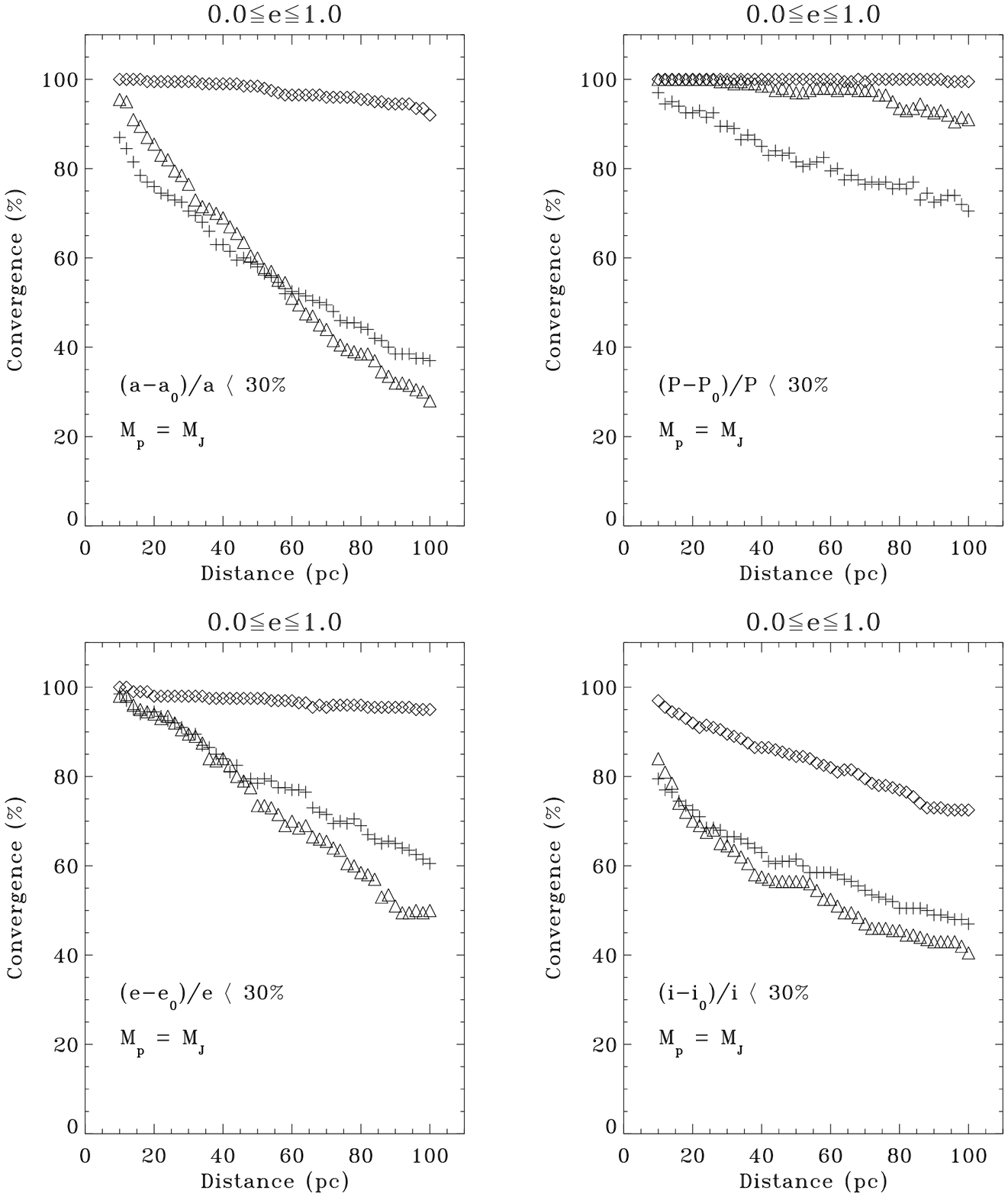}
\caption{{\em Same as Figure~\ref{conv}, for convergence to a 30 per cent
accuracy level.}}
\label{conv2}
\end{figure}

\begin{figure}
\vspace{11cm}
\includegraphics{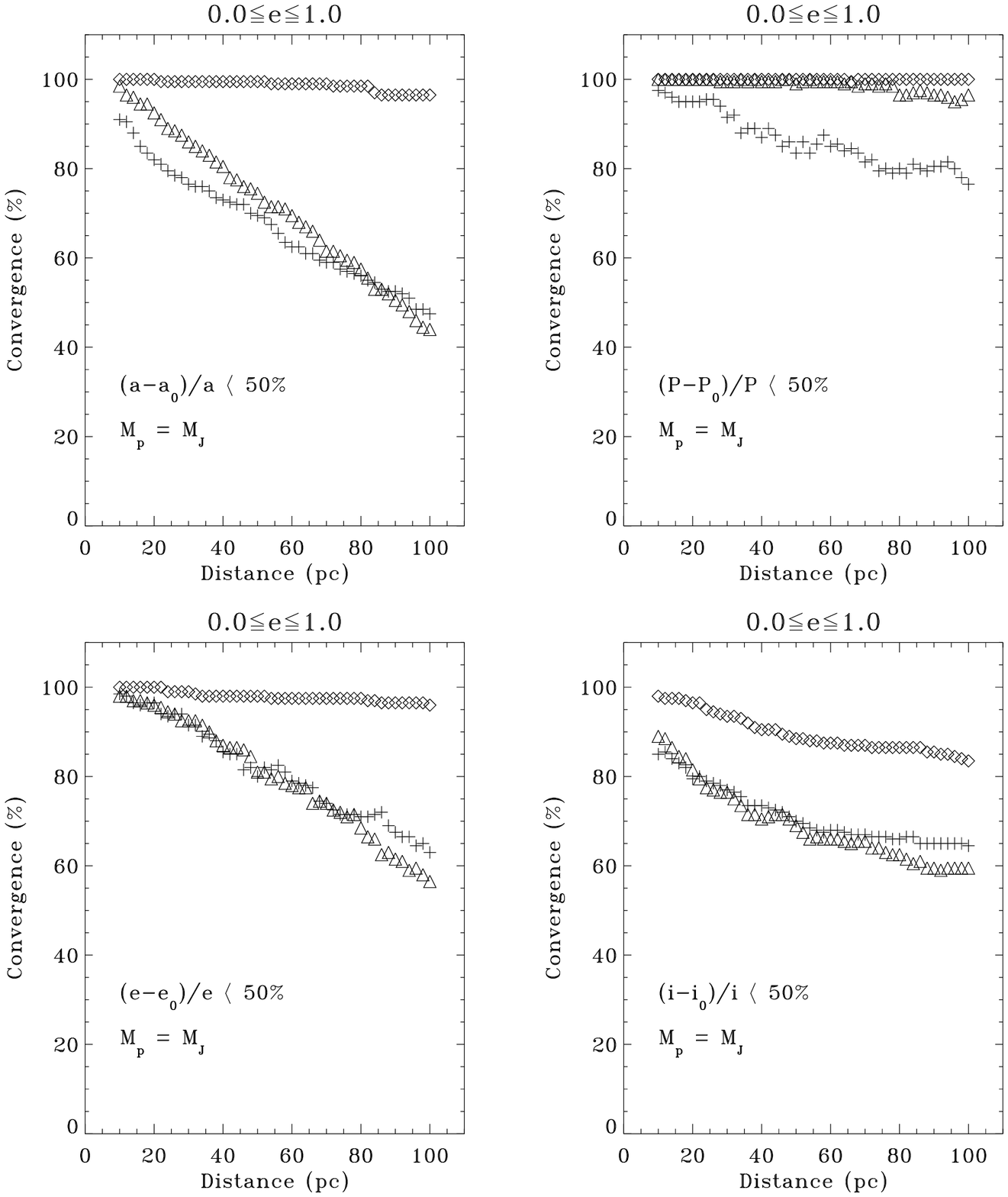}
\caption{{\em Same as Figure~\ref{conv}, but for convergence to a 50 per cent
accuracy level.}}
\label{conv3}
\end{figure}

\begin{figure}
\vspace{15cm} \includegraphics{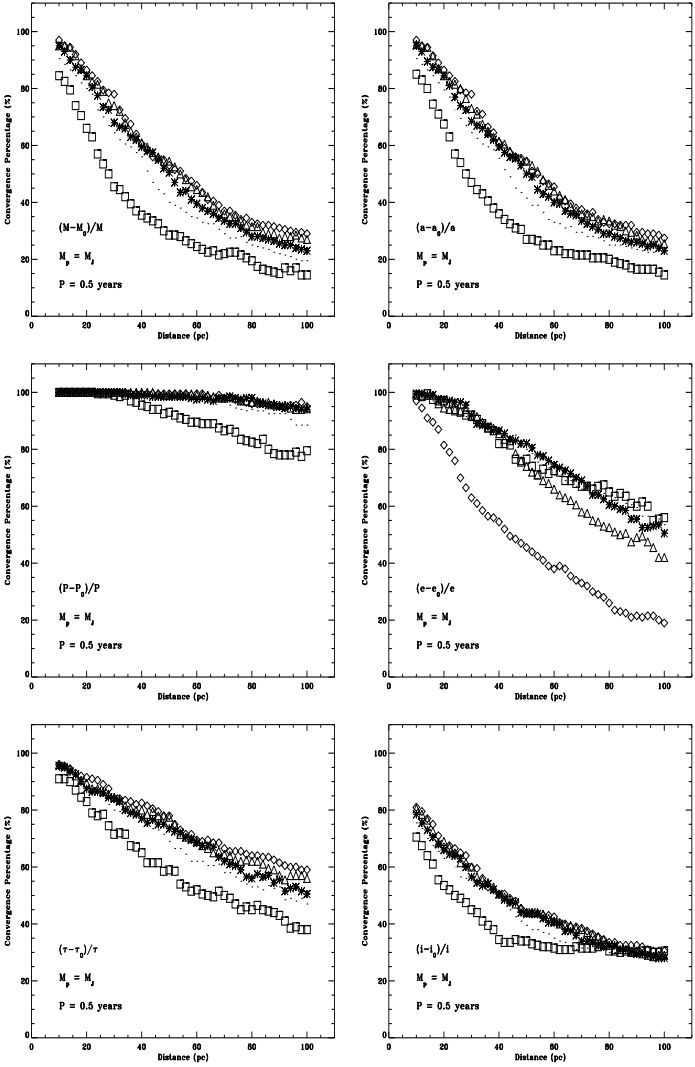} \caption{{\em The effect of eccentricity:
the plots show the 20 per cent accuracy level in cases of
Sun-Jupiter systems with orbital period P = 0.5 years and $e$ set
at: 0.1 (diamonds), 0.3 (triangles), 0.5 (stars), 0.7 (points),
0.9 (squares).}} \label{ecc}
\end{figure}

\begin{figure}
\vspace{15cm} \includegraphics{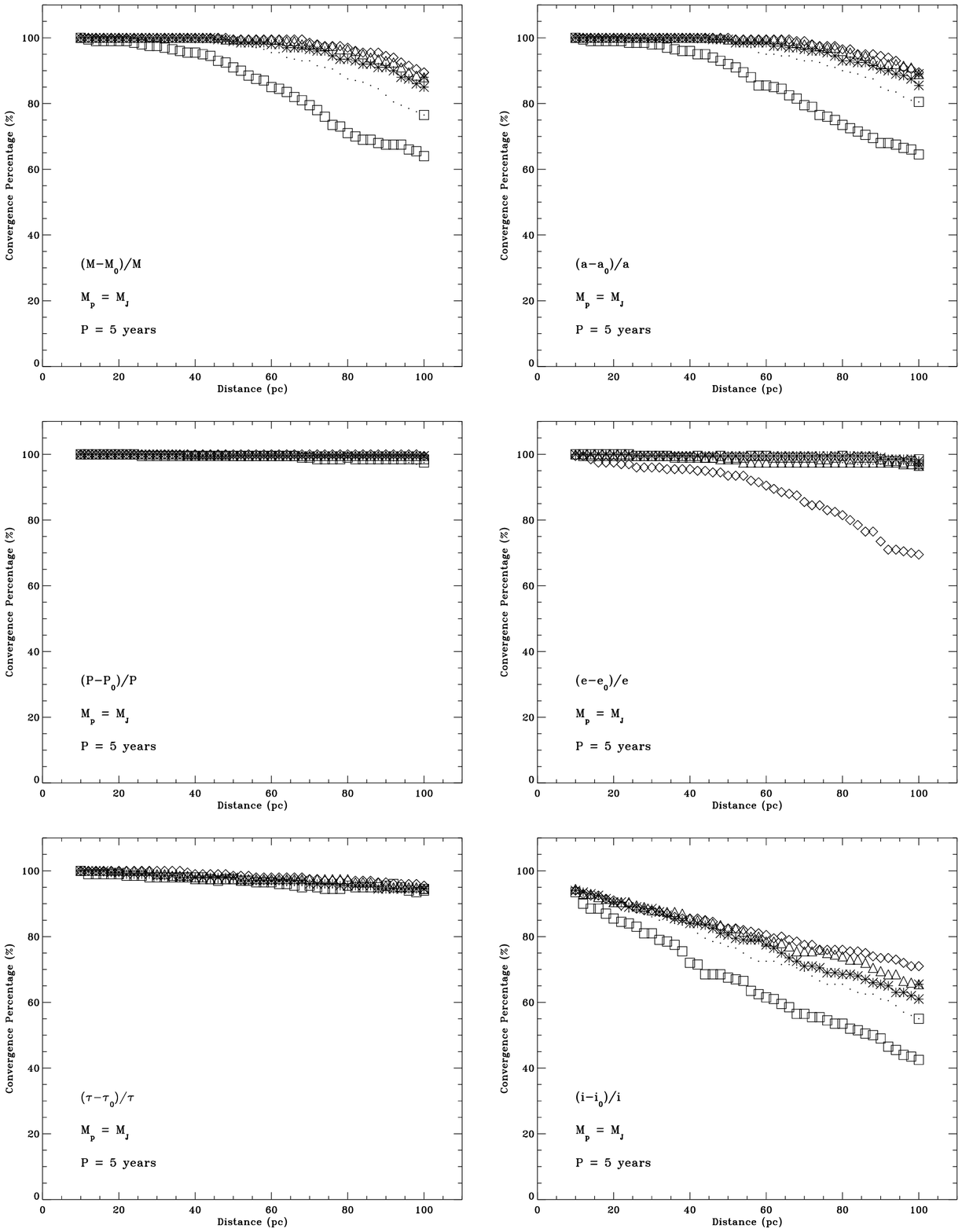} \caption{{\em Same as Figure~\ref{ecc},
for Sun-Jupiter systems with P = 5 years.}} \label{ecc2}
\end{figure}

\begin{figure}
\vspace{15cm} \includegraphics{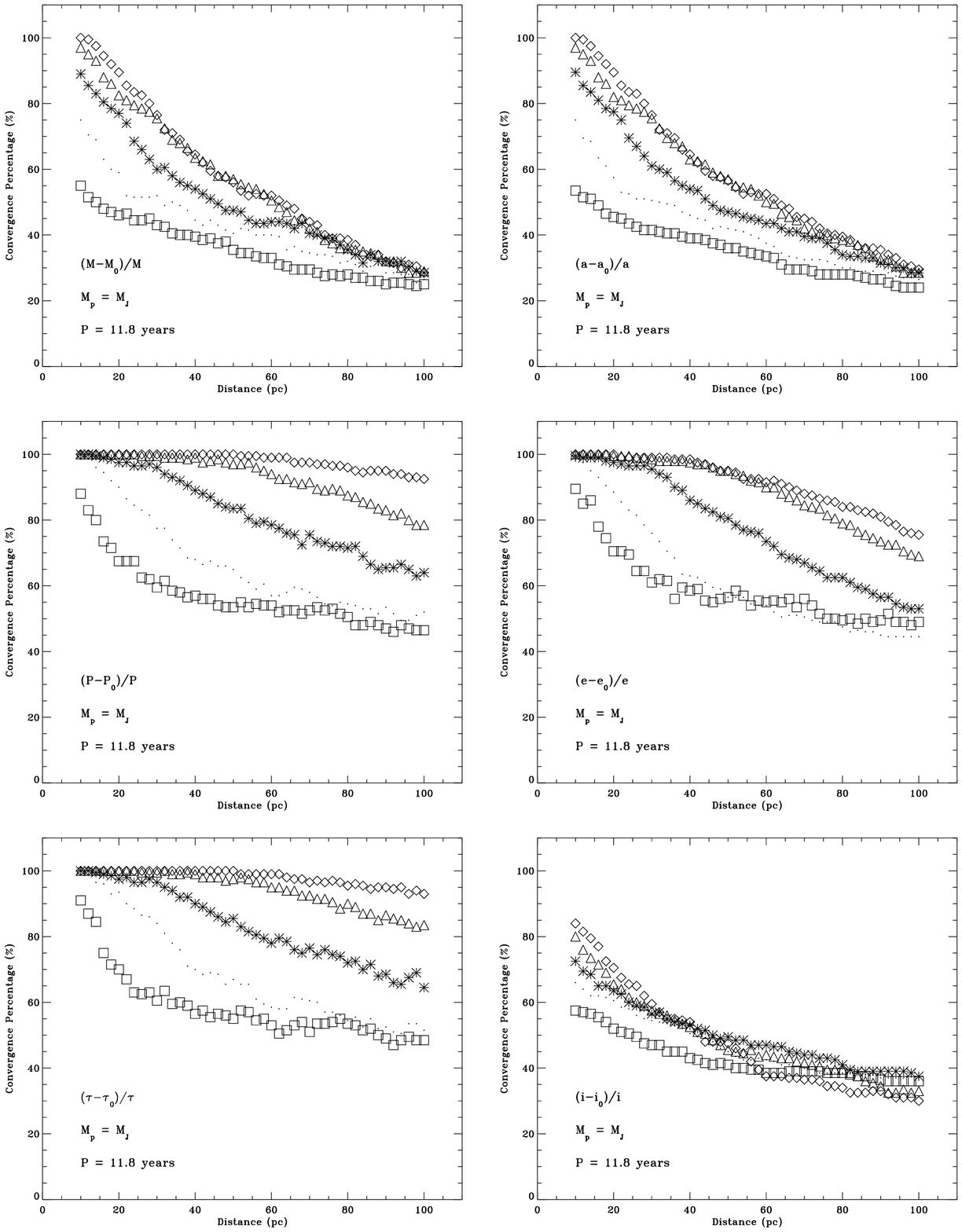} \caption{{\em Same as Figure~\ref{ecc},
for Sun-Jupiter systems with P = 11.8 years.}} \label{ecc3}
\end{figure}

Similarly to what we did in section~\ref{detect}, 
we look at the accuracy of the results
primarily as function of distance and orbital period.  We consider
systems with Sun-Jupiter masses and three values of the period: 0.5
years, to test {\it GAIA}'s ability to cope with poorly-sampled motion;
5 years, a near-ideal case where the orbital period is as long as the
mission; and 11.8 years, the true period of the Jupiter-Sun system,
which stretches {\it GAIA}'s ability to solve long-period orbits.
For each case, we generate 200 systems randomly placed on the sky,
with randomly distributed values of the other orbital parameters.
These 200 systems are then placed all at the same distance,
and the simulation is repeated for distance values ranging from
from 10 to 100~pc.

A fit to the simulated observations is made directly at the GC level,
taking into consideration a theoretical model in which {\it GAIA}'s
unidimensional measurements are expressed as functions of the 5
astrometric parameters plus the 7 orbital elements of the keplerian
orbit: for the latter we adopted a slightly linearized analytic form
where semi-major axis, inclination, periastron longitude and position
angle of the ascending node are combined in the four Thiele-Innes elements,
while only orbital period, eccentricity and periastron epoch are
left unchanged.

The first solution of the linearized system is thus found
with respect to the initial guess. Then, the linearization at step $k$ 
is updated with respect 
to the solution obtained at step $k-1$ and the process repeated until the  
differential corrections $\delta a_{i,k}$  to each of the 12 
parameters satisfy the relation: 
$\left|\delta a_{i,k}/a_{i,k-1}\right| < 10^{-6}$, where $i = 1,2,\dots,12$ 
and $a_{i,k-1}$ are the parameters adjusted at the previous step.
  We then record what fraction of the final
values for each parameter falls within a certain fractional error
of the true value: this we call the ``convergence probability'',
which is a function of the distance of the system, the period, and the
desired fractional error.

We evaluate the convergence probability for the parameters which are most
likely to affect the efficiency of the reconstruction of the orbit and of the
mass determination, namely semimajor axis $a$, period $P$, inclination $i$,
and eccentricity $e$. For each parameter, we consider fractional errors
of 10, 30, and 50 per cent.

The results are shown in Figures~\ref{conv} to~\ref{conv3}. Points
of different shapes correspond to different periods: triangles for
0.5 years, diamonds for 5 years, and crosses for 11.8 years.
Figure~\ref{conv} shows the probability of convergence within 10
per cent of the true values; Figure~\ref{conv2} to within 30 per
cent; and Figure~\ref{conv3} to within 50 per cent. In each
Figure, different panels correspond to different quantities.

As expected, the 5 year-period case is the best of the three.  The 11.8
year period, although it corresponds to a larger signal (as the
semimajor axis, and thus $ \alpha $, increases with period), suffers
from the incomplete sampling of the orbit during the mission life-time,
while the short-period case suffers from both the smaller signal
amplitude and a (generally) non-optimal timing of the observations.

The periodicity of the signal is the characteristic which can be
evaluated with the best accuracy: $P$ is the only element which is always
estimated with high accuracy (better than 10\ per cent) throughout the
ranges covered by our simulations.  We also note that the semimajor
axis $ a $ and the inclination $ i $ behave similarly to each other.
This may be in part a consequence of the use of the Thiele-Innes 
representation, since both parameters are obtained from combination of 
the Thiele-Innes elements.

Somewhat surprisingly, orbital eccentricity---which in the previous
simulation is assumed to be distributed uniformly from 0 to 1---has a
very significant effect on the quality of the estimated orbital
parameters, including the mass of the planet.  We illustrate this in
Figures~\ref{ecc} through~\ref{ecc3}, where the different symbols now
refer to different orbital eccentricities, and the convergence
probability is given for a fixed fractional error of 20 per cent.  Different
figures correspond to different orbital periods.  Generally,
high eccentricities deteriorate convergence percentages for all
orbital parameters. 
The reason is that the regularly-spaced observations of a survey
satellite, such as {\it GAIA}, may be ill-suited to sample an orbital
motion with large velocity variations, as happens for high eccentricities.  The
effect is especially prominent for the long-period case, $ P = 11.8 $~years,
for such orbits will often {\it never} be observed during the periastron.

\vspace {20pt}

Our main results can thus be summarized as follows:

\begin{itemize}

\item[--] For a uniform eccentricity distribution in the range $0\le e\le 1$,
then more than a half of all existing Jupiter-mass planets orbiting
around solar-type stars with periods comparable to the mission lifetime
can be detected, and their masses and orbital parameters can be
estimated to an accuracy of 10 per cent up to distances of about 100 pc.
From the completeness limit of the {\it HIPPARCOS} catalog ($V\sim 7.5-8$ mag)
and considering spectral types no later than G5, we estimate 20\,000, 
65\,000, and 150\,000 stars to 100, 150, and 200 pc respectively. 
In this estimate the contribution from early type stars and giants is 
negligible. Note that these numbers increase significantly when considering 
spectral types earlier than K5: calculations based on current Galaxy models 
predict $\gsim$ 300\,000 stars within 200 pc from the Sun~\cite{latt99}.

\item[--] For a fractional error no worse than 20 per cent, the distance limit
is reduced to 50 pc for short-period and long-period orbits 
(0.5 and 11.8 years);

\item[--] Low-eccentricity systems are easier to detect and solve: this
could indicate a potential limit in {\it GAIA}'s capability of
reconstructing the precise behaviour of the mass function of low mass
companions to solar-type stars in the neighborhood of our Sun, as to our
knowledge first observational data seem to indicate than brown dwarfs
are more likely to revolve around the parent stars on significantly
eccentric orbits~\cite{black}.  
On the other hand, low orbital eccentricities may
be prevalent for longer periods, thus enhancing {\it GAIA}'s capabilities
for such systems.

\end{itemize}

\subsection{47 Uma, 70 Vir, and 51 Peg}

\begin{figure}
\vspace{9cm}
\includegraphics{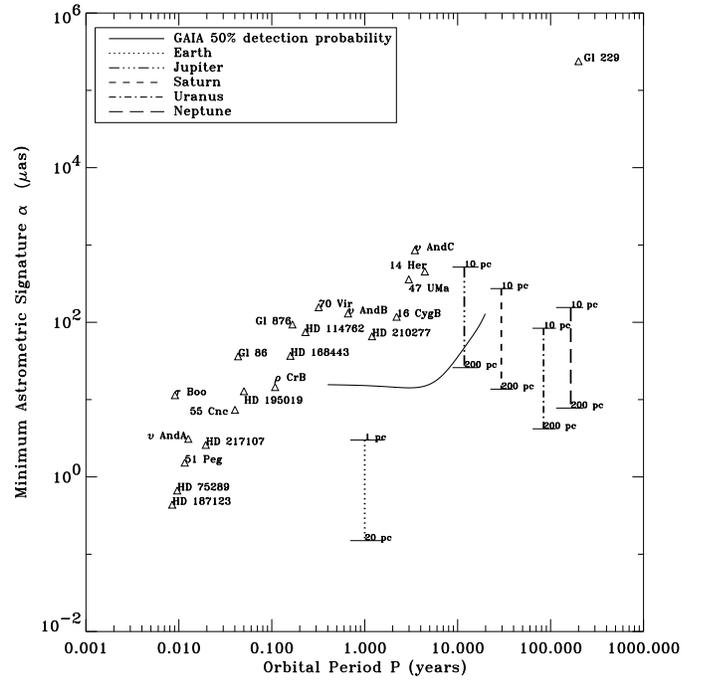}
\caption{\em Comparison between GAIA 50 per cent probability curve 
(solid line) and the minimum astrometric signature ($\alpha$)
induced on the parent star by presently known planetary mass objects
 ($0.5\, M_J\le M\le 11\, M_J$) in the neighborhood of our sun, as function 
of orbital period (log-log plot). The dahsed, dotted, and 
dashed-dotted lines represent 
the astrometric signatures induced on a solar-like star by the 
major solar system bodies, at increasing 
distance from the observer}
\label{neigh}
\end{figure}

We now consider more specifically how the known planetary systems fall
within {\it GAIA}'s capabilities. {\it Figure~\ref{neigh} shows period and
astrometric signature for the planets detected to date by radial
velocity measurements as compared to {\it GAIA}'s 50 per cent 
iso-probability curve}. This is the {\it minimum} signature, corresponding
to an orbital inclination of 90$^\circ $; the radial velocity method cannot
generally remove the degeneracy between planetary mass and inclination.
Although all systems are very close to the Sun ($d\lsim 40$~pc),
several planets have very small signatures due to their short
orbital period (and thus small semi-major axis, by Kepler's third law).  Some
of the shortest-period objects, i.e, the 51~Peg-class planets,
may prove difficult {\it to detect and measure} with {\it GAIA}. On the
other hand, planets with signature $\ge 100 \muas $ will 
be easily detected, and
their orbital elements can be found with high accuracy by {\it GAIA}.

The considerations and results obtained in the previous sections from a
statistical viewpoint are obviously preliminary to the much more
difficult task of the development of self-consistent detection and
orbital parameters estimation algorithms devoted to the {\it investigation of 
individual objects}, in which a more complete and realistic error
model can be taken into account: to this end detailed system studies are
in progress and will be presented elsewhere.  It is nevertheless of
interest a first glance to how well {\it GAIA} will behave once at work:
thus, to test further {\it GAIA}'s capabilities of detecting
periodic oscillations and signatures due to planetary companions around
stars in the neighborhood of our solar system, we have concentrated our
attention on the results obtained with radial velocity techniques, for
what concerns three of {\it the star-planet systems known to date}: 47 UMa,
70 Vir and 51 Peg.  All these stars are nearby and very similar to our
Sun.
\begin{figure*}
\vspace{14cm}
\includegraphics{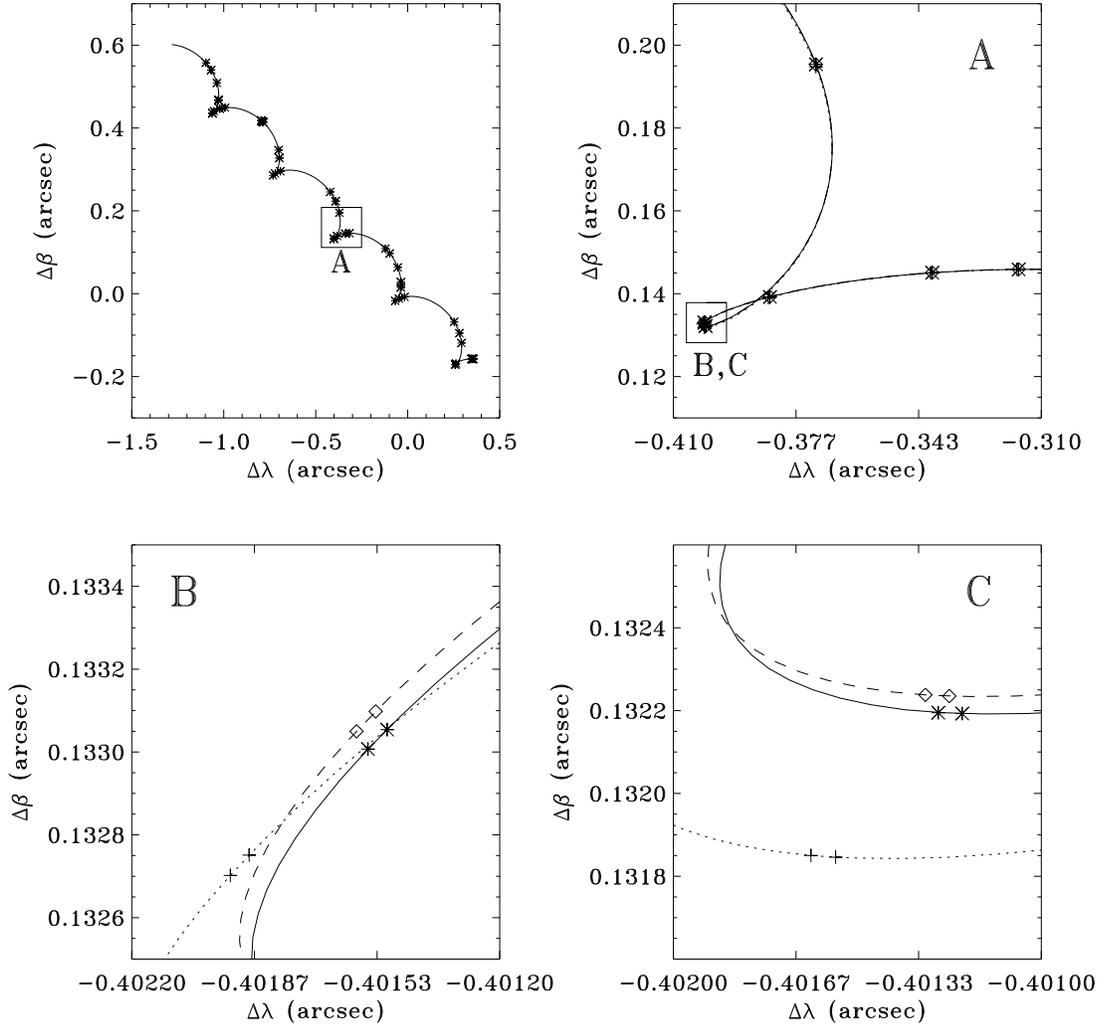}
\caption{{\em Planetary perturbations of 47 UMa as ``seen'' by {\it GAIA} for an orbital
inclination of 45$^\circ$.
The top left panel shows the path (solid line) of the star over a 5-year
period. The $*$ represent the {\em true} location of the star at the time
of the simulated {\it GAIA} observations. The upper right panel 
zooms on the $0.1^{\prime\prime}\times 0.1^{\prime\prime}$ 
region marked with a small square in the first panel. 
As for the two bottom graphs, 
they each represent a different 1-mas$^2$ enlargement of the top right panel.
Again, the solid line represents the true (simulated) path of the parent
star of the system. A single-star fit for 47 UMa produces the
dotted path with crosses marking the calculated positions at the same
observing times as before. The dashed line shows the recomputed positions
after fitting the observations with a 10-parameter model (five astrometric
parameters plus the four Thiele-Innes elements and the orbital period)}}
\label{47uma}
\end{figure*}

According to the spectroscopic measurements~\cite{uma,vir,mayo},
around these three stars orbit planetary bodies
of minimum mass, respectively,
$\sim$2.46 $M_{\rm J}$, $\sim$6.50 $M_{\rm J}$, and $\sim$0.5 $M_{\rm J}$,
with orbital periods of about 3 years, 4 months, 4 days.  This
translates in the following astrometric signatures:

\begin{center}
$\alpha$ (47 UMa) $\ge$ 361.88 $\mu$as
\end{center}
\begin{center}
$\alpha$ (70 Vir) $\ge$ 167.93 $\mu$as
\end{center}
\begin{center}
$\alpha$ (51 Peg) $\ge$ 1.66 $\mu$as
\end{center}

Because of its extremely short period, the astrometric perturbation in
the case of the 51 Peg system is so small that 
reliable estimates of its astrometric orbit with {\it GAIA} will be highly 
difficult. 
On the other hand, the detection of the signatures induced on 47 UMa and 70
Vir will be extremely easy for a {\it GAIA}-like satellite.

We have generated 100 simulated planetary systems on the celestial
sphere, respectively identical to 47 UMa and 70 Vir, assuming perfectly
circular orbits and stellar masses $M$ = $M_\odot$.  The unknown orbital
inclination was chosen to be $i=45^\circ$.

First of all, we applied the $\chi^2$ test to the simulated
observational data, with a conservative single-observation error
$\sigma_\psi=10$ {\muas}. The detection probabilities derived are, as
expected, always about 100 per cent, given the signal-to-noise ratio to
be always in the regime $S/N \gg 1$.
\begin{table*}
\noindent\begin{minipage}{145mm} \caption{{\em Results of fitting
{\it GAIA} observations of 47 UMa: the true orbital period
simulated is P$_{\rm true}$ = 2.98 yr, the star's parallax is
$\pi_{\rm true}$ = 71.03 mas. The estimated values are averages
over 200 simulations per orbital inclination bin. The RMS errors
are derived from direct comparison to the true values.}}
\label{47tab}
\begin{tabular}{@{}ccccccc}
\hline & & & & & & \\ $i$&$a_{{\rm true}}$ ($\mu$as) & $a_{{\rm
fit}}$ ($\mu$as)&P$_{{\rm fit}}$ (years)& $\pi_{{\rm fit}}$ (mas)&
M$_{{\rm p,true}}$ (M$_{{\rm J}}$)&M$_{{\rm p,fit}}$ (M$_{{\rm
J}}$)\\ & & & & & & \\ \hline
$20^\circ$&1058.08&$1060.92\pm 15.15$&$2.951\pm 0.125$&
$71.0377\pm 0.0026$&7.19&$7.27\pm 0.24$\\
$30^\circ$&723.77&$725.56\pm 14.45$&$2.956\pm 0.142$&$71.0377\pm
0.0029$& 4.92&$4.97\pm 0.19$\\
$40^\circ$&562.99&$565.62\pm 14.97$&$2.955\pm 0.119$&$71.0381\pm
0.0027$& 3.83&$3.87\pm 0.15$\\
$50^\circ$&472.41&$470.75\pm 14.82$&$2.958\pm 0.158$&$71.0377\pm
0.0030$& 3.21&$3.22\pm 0.15$\\
$60^\circ$&417.87&$419.63\pm 17.07$&$2.959\pm 0.179$&$71.0372\pm
0.0028$& 2.84&$2.87\pm 0.16$\\
$70^\circ$&385.11&$384.36\pm 14.48$&$2.973\pm 0.112$&$71.0378\pm
0.0026$& 2.62&$2.62\pm 0.12$\\
$80^\circ$&367.47&$367.93\pm 16.38$&$2.962\pm 0.157$&$71.0381\pm
0.0025$& 2.50&$2.51\pm 0.14$\\
$90^\circ$&361.88&$362.64\pm 15.17$&$2.930\pm 0.159$&$71.0377\pm
0.0023$& 2.46&$2.50\pm 0.14$\\

& & & & & & \\
\end{tabular}
\end{minipage}\par
\end{table*}
\begin{table*}
\noindent\begin{minipage}{145mm} \caption{{\em Results of fitting
{\it GAIA} observations of 70 Vir: the true orbital period
simulated is P$_{\rm true}$ = 0.32 yr, the star's parallax is
$\pi_{\rm true}$ = 55.22 mas.}} \label{70tab}
\begin{tabular}{@{}ccccccc}
\hline & & & & & & \\ $i$&$a_{{\rm true}}$ ($\mu$as) & $a_{{\rm
fit}}$ ($\mu$as)&P$_{{\rm fit}}$ (years)& $\pi_{{\rm fit}}$ (mas)&
M$_{{\rm p,true}}$ (M$_{{\rm J}}$)&M$_{{\rm p,fit}}$ (M$_{{\rm
J}}$)\\ & & & & & & \\ \hline
$10^\circ$&967.05&$974.56\pm 15.83$&$0.3198\pm 0.0015$&
$55.2216\pm 0.0033$&37.43&$37.73\pm 0.62$\\
$20^\circ$&490.98&$495.70\pm 15.54$&$0.3199\pm 0.0016$&
$55.2211\pm 0.0027$&19.00&$19.19\pm 0.60$\\
$30^\circ$&335.85&$339.64\pm 14.36$&$0.3201\pm 0.0021$&
$55.2208\pm 0.0030$&13.00&$13.14\pm 0.56$\\
$40^\circ$&261.25&$265.61\pm 15.37$&$0.3201\pm 0.0023$&
$55.2215\pm 0.0030$&10.11&$10.28\pm 0.60$\\
$50^\circ$&219.21&$218.48\pm 15.39$&$0.3198\pm 0.0023$&
$55.2209\pm 0.0030$&8.49&$8.46\pm 0.60$\\
$60^\circ$&193.91&$195.89\pm 17.03$&$0.3199\pm 0.0029$&
$55.2205\pm 0.0028$&7.51&$7.58\pm 0.66$\\
$70^\circ$&178.70&$181.71\pm 14.71$&$0.3198\pm 0.0032$&
$55.2215\pm 0.0025$&6.92&$7.03\pm 0.57$\\
$80^\circ$&170.52&$172.00\pm 14.23$&$0.3197\pm 0.0030$&
$55.2211\pm 0.0026$&6.60&$6.66\pm 0.55$\\
$90^\circ$&167.93&$169.69\pm 14.31$&$0.3198\pm 0.0042$&
$55.2210\pm 0.0029$&6.50&$6.57\pm 0.56$\\ & & & & & & \\
\end{tabular}
\end{minipage}\par
\end{table*}

In order to obtain a visual indication of the quality of the
reconstruction, we then compared the apparent motion of the star
with the motion obtained with a single star fit and that obtained
by a fit with the fitted parameters for the planet.
Figure~\ref{47uma} shows the differences in 47 UMa's path on the
sphere, when calculated starting from: a) the true values of
$\mu_\lambda$, $\mu_\beta$, $\pi$, adopted in the simulation
(asterisks, solid line); b) the values of $\mu_\lambda$,
$\mu_\beta$, $\pi$, obtained after fitting the observations with a
single star model (crosses, dotted line); and c) the values of
$\mu_\lambda$, $\mu_\beta$, $\pi$ and of the orbital parameters
derived after fitting the observations with a 10 parameter model
reproducing the standard motion plus the Keplerian circular motion
induced on the observed star by the presence of the planet
(diamonds, dashed line).

The annual proper motion of 47 UMa is very large, compared to the
magnitude of the astrometric signature produced by the
gravitational influence of its planetary companion: in ecliptic
coordinates, $\mu_\lambda$ = -0.283 arcsec/year, $\mu_\beta$ =
0.152 arcsec/year, while $\alpha \sim 0.5$ mas for an orbital
inclination of $45^\circ$. The top left panel shows the full range
of motion of 47 UMa over five years; the looping motion is the
parallactic eclipse, and the effect of the planet is essentially
undetectable on this scale. In order to see the difference in the
residuals, we zoom first on region A marked by a small square 
in the top left panel, then we identify in the upper right panel two 
1-mas$^2$ regions (B and C), representing each a 1000x
enlargement of a small fraction of the motion (bottom panels).
The true and reconstructed motions are very close, consistent with
the $10 \mu$as single-measurement error of the observations used
to derive the orbital parameters, while on this scale the orbit
obtained with the single-star assumptions (dotted) is clearly very
different from the actual motion.

Third, we derive the uncertainties in the fitted values of the
important physical parameters of the two systems, and especially
the companion masses.  We derive the mass from the fitted orbital
parameters via the mass function:
\begin{equation}
\frac{M_p^3}{(M_s + M_p)^2} = \frac{a_s^3}{\pi^3}\frac{1}{P^2}
\end{equation}
 where again $M_p$ and $M_s$ are the planetary and stellar masses
in solar masses, $\pi$ the parallax, $ P$ the orbital period in
years, and $a_s$ the semi-major axis of the parent star, expressed in
arcsec.

Assuming $M_p \ll M_s$,  we obtain:
\begin{equation}
 M_p \simeq
\left(\frac{a_s^3}{\pi^3}\frac{M_s^2}{P^2}\right)^{1/3}
\label{t:mass}
\end{equation}

Fitting the simulated observations with a 10 parameter model we
can derive estimates of $\pi$ and $P$ directly, while $a_s$ can
be obtained as a combination of the approximated four Thiele-Innes
elements. Provided that one gets informations about the  mass of
the primary, for example thanks to spectroscopy, it is possible to
calculate $M_{p}$.

In our case we have, for simplicity, supposed to know $M_{47 UMa}$ and
$M_{70 Vir}$ to be exactly equal to $M_\odot$.

Tables~\ref{47tab} and~\ref{70tab} show the approximated values of
$M_{p}$ in the two cases, starting from the estimated values of
$P$, $\pi$ and $a_s$ obtained during the fit, as function of
orbital inclination.  The Tables stop at $i=20^\circ$ for 47 UMa
and $i=10^\circ$ for 70 Vir, taking into consideration
 the upper limits on their masses found by
Perryman et al~\shortcite{perry} using {\it HIPPARCOS} data. The
{\it HIPPARCOS} measurements, as a matter of fact, agreed with the
stars to be single, as no evident signature greater than the
nominal error ($\sim$1 mas) was revealed. This implies the
following upper limits: $M_{p}$$\le$ 7 $M_{\rm J}$, for 47 UMa,
$M_{p}$$\le$ 38 $M_{\rm J}$, for 70 Vir, which again means,
considering the results obtained spectroscopically, choosing a
minimum orbital inclination as in the tables. In both cases, it is
clearly evident the high accuracy with which the relevant orbital
parameters and the planet's mass are recovered: thanks to the fact
that $S/N \gg 1$, the fit to {\it GAIA}'s simulated observations
of 47 UMa and 70 Vir is very satisfactory; the RMS errors
between fitted and nominal values of the various parameters are
always well within 10 per cent.

Such results confirm that {\it GAIA} could reveal itself a very
powerful instrument to investigate many of the known candidate
planetary systems, with the exception of very short-period
systems, such as 51 Peg: provided the signal-to-noise ratio is
sufficiently high and the period not too short, our simulations
show that such systems will be easily detected,  and their
orbital parameters accurately determined by {\it GAIA}, at least
if such systems are effectively simple, as the first spectroscopic
measurements would suggest.  In case the signatures produced
cannot be interpreted as due to the presence of only one planet,
this would introduce many interpretative complications in the
signals observed: for planetary systems resembling our
own, detection should be little affected and residuals analysis might
reveal hints of the influence of the outer planets
($\alpha$ for a Saturn-like companion is $\sim$ 60 {\muas}
at 10 pc but its period is $\sim$ 6 times the mission lifetime).
However, reliable orbital fitting would probably 
be restricted to the main component.
On the other hand,
multiple orbital fitting to good accuracy should be possible for
systems composed of giant planets of comparable mass, orbiting 
with periods within the interval $\sim$0.5-5 yr, and
yielding astrometric S/N $\gsim$ 10 , like it would be
the case for the two outermost planets of the recently
discovered system $\upsilon$ And~\cite{butl}.

The capability of detecting and measuring 
multiple planets
with {\it GAIA} will become matter of future simulations and
system studies.

\section{Summary and conclusions}

In this work we have given the first quantitative evaluation of the
detectability horizon of single extra-solar giant planets around single normal 
stars in the neighborhood of our solar system for the global astrometry 
mission {\it GAIA}. Complete simulations, comprehensive of observations of 
star-planet systems and
successive statistical analysis of the simulated data, have yielded the
following results:

\begin {itemize}

\item [a)] it will be possible to detect more than 50 per cent of all
{\em Jupiter-like} planets (orbital period $P = 11.8$ years) orbiting
solar-type stars within 100 pc; {\em Jupiter-size} planets, with shorter
orbital periods, will be detectable up to 200 pc, with similar
probabilities;

\item [b)] for {\em true Sun-Jupiter systems} it will be possible to
determine the full set of orbital parameters and to derive accurate
estimates of the masses up to distances of order of 50 pc, value which
doubles if we consider the range of periods in the vicinity of the
mission lifetime;

\item [c)] simulated observations of a selection of the actually known
extra-solar planets, discovered by means of spectroscopic measurements,
provide a meaningful estimate of the uncertainty with which masses and
orbital elements can be determined for the known star-planet systems and
for a substantial fraction of those that will be found within the
context of such a global astrometry mission.
Although preliminary, our results indicate that these systems will be
easy to discover and their orbital parameters will be accurately
determined with {\it GAIA}, except very short-period systems such as 51
Peg.

\end {itemize}

Hence, our results indicate that: $1)$ 
{\it GAIA} would monitor all of the hundreds of 
thousand F-G-K stars (i.e., whose masses are within a factor $\sim$ 1.5 
that of the Sun) up to a distance of $\sim$ 200 pc from the Sun, in
search for astrometric signatures due to the presence of giant planets 
($M\simeq M_J$) with orbital periods up to Jupiter's; 
$2)$ a significant fraction of the detected planets would 
have the main orbital parameters (semi-major axis, period, eccentricity, 
inclination) measured to better than 30 per cent accuracy. 

Therefore, the {\it GAIA} survey would uniquely complement the expectations 
from other ongoing and planned spectroscopic and astrometric planet searches, 
both from ground and in space, thus helping with the creation of
the fundamental testing ground on which to measure the validity of 
actual theoretical models of formation
and evolution. {\it GAIA}'s discovery potential might have significant impact 
on our knowledge of the distribution laws of the most relevant orbital 
parameters, and it would contribute
to determine the frequency of planetary systems themselves in the solar
neighborhood and, by extrapolation, in the whole Galaxy.  A vast
all-sky astrometric survey would help understand peculiar 
characteristics of these systems, e.g.,
whether giant planets lying in the outer regions are common:
such planetary scenarios may be worth further
investigation, as, according to present theoretical models, this could
indicate presence of low mass planets in the inner regions, possibly in
the parent stars' habitable zones. The monitoring of hundreds of
thousands stars directly implies the chance to investigate objects
belonging to a wide range of spectral types, thus providing the important
observational material~\cite{boss} to decide whether giant planets are
more likely to form by means of gravitational instability in disks (once
they are found more often around young
stars~\cite{kui,came,bode}), or by means of accretion of
planetesimals (once they are found more often around old
stars~\cite{poll,lissa,polla}). The high precision
global astrometric measurements will estimate the inclination $i$ of the
orbital planes for the majority of the presently known planetary systems
and for a large fraction of those that will be eventually discovered: it
will then be possible to provide unambiguous mass estimations of such
dark companions, reducing significantly the uncertainty on the mass
range in the transition region from brown dwarfs to giant planets.

\section*{Acknowledgments}

We wish to offer our special thanks to M.A.C.  Perryman for lending
initial impetus and continuing support to this investigation.  Over the
course of this work, we have benefited from discussions with numerous
colleagues, and especially Gerry Gilmore, David Latham,
Lennart Lindegren, Robert Reasenberg and Stuart Shaklan. 
Also, we wish to thank the referee for her/his careful comments 
which helped us improve the original manuscript.
All four authors gratefully acknowledge partial financial support from
the Italian Space Agency, under contract ASI/ARS-96-77.

\label{lastpage}
\end{document}